\documentclass[twocolumn,aps,prd]{revtex4-2}
\usepackage{graphicx}
\usepackage{bm}
\usepackage{relsize}
\usepackage{xfrac}
\begin{document}
\newcommand*{\bi}{\bibitem}
\newcommand*{\ea}{\textit{et al.}}
\newcommand*{\eg}{\textit{e.g.}}
\newcommand*{\zpc}[3]{Z.~Phys.~C \textbf{#1}, #2 (#3)}
\newcommand*{\plb}[3]{Phys.~Lett.~B \textbf{#1}, #2 (#3)}
\newcommand*{\phrd}[3]{Phys.~Rev.~D~\textbf{#1}, #2 (#3)}
\newcommand*{\ibid}[3]{\textit{ ibid.} \textbf{#1}, #2 (#3)}
\newcommand*{\epjc}[3]{Eur. Phys. J. C \textbf{#1}, #2 (#3)}
\newcommand*{\ra}{\rightarrow}
\newcommand*{\pippim}{\pi^+\pi^-}
\newcommand*{\rf}[1]{(\ref{#1})}
\newcommand*{\be}{\begin{equation}}
\newcommand*{\ee}{\end{equation}}
\newcommand*{\die}{e^+e^-}
\newcommand*{\dik}{\mathrm{K}^+\mathrm{K}^-}
\newcommand*{\jj}{\mathrm i}
\newcommand*{\cndf}{\chi^2/\mathrm{NDF}}
\newcommand*{\minuit}{\texttt{MINUIT}~}
\def\babar{\mbox{\slshape B\kern-0.1em{\smaller A}\kern-0.1em
B\kern-0.1em{\smaller A\kern-0.2em R}}}

\title{
Can the X(1750) be a \bm{$\phi$}(1750)?}
\author{Peter Lichard}   
\affiliation{Institute of Physics and Research Centre for Computational 
Physics and Data Processing, Silesian University in Opava, 746 01 Opava, 
Czech Republic}

\begin{abstract}
We present the arguments that the X(1750) has the isospin $I=0$, and as such, 
it belongs to the $\phi$ family. We base our 
argumentation on the X(1750) appearing 
in the $\die\ra\dik$ process as a narrow resonance and not appearing 
in the $\die\ra\pi^+\pi^-$ process at all.
\end{abstract}
\date{\today}
\maketitle

The X(1750) resonance was discovered in 1981 at the CERN Omega Spectrometer 
\cite{aston81f} with the mass of (1748$\pm$11)~MeV and the width of
(80$\pm$30)~MeV \cite{pdgwrong}. Other photoproduction experiments \cite{gamma} 
also confirmed its existence. The BESIII Collaboration \cite{ablikim20f} 
has recently made an important observation that X(1750) appears together 
with $\phi(1680)$ in decays of $\psi(2S)$.

The affinity of X(1750) for kaons is obvious from all experiments.
With its well-established quantum numbers $J^{PC}\!=1^{--}$
the X(1750) should reveal itself in the direct channel of the
electron-positron annihilation into $\dik$. 

There are only two data sets on electron-positron annihilation into 
charged-kaon pairs in the cms energy region from the threshold to 2~GeV:
(i) 2013 data by \babar~Collaboration \cite{babar2013KpKm}, and (ii)
data taken at the SND detector at the VEPP-2000 $\die$ collider in
Novosibirsk published in 2016 \cite{snd2016KpKm}. Data from
the CMD-3 Collaboration advertised in 2020 \cite{ivanov2020} have not yet
appeared in a final form. In our quest to determine the isospin of 
the X(1730) resonance, we will use very precise \babar~data
\cite{babar2013KpKm} that cover the energies up to 5~GeV. We will 
limit ourselves to energies below 2.2~GeV and consider only statistical 
errors.

\begin{figure}[b]
\includegraphics[width=0.39\textwidth,height=0.11\textwidth]{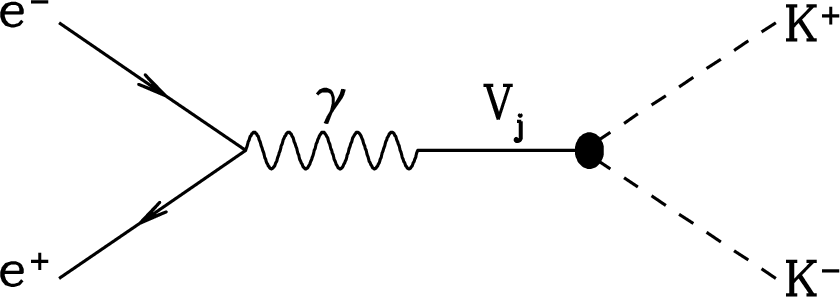}
\caption{\label{fig:ee2KpKm}Feynman diagram of the VMD model.}
\end{figure}

For the description of the electron-positron annihilation into a $\dik$ pair,
 we used a vector-meson-dominance (VMD) model with $n$ intermediate vector 
resonances based on the Feynman diagram depicted in the 
Fig.~\ref{fig:ee2KpKm}.
It provides the cross-section formula 
\be
\label{sigma}
\!\sigma(s)=\frac{\pi\alpha^2}{3s}\left(1\!-\!\frac{4m_\mathrm{K}^2}{s}
\right)^{\!\!3/2}
\left|\sum_{j=1}^n
\frac{R_je^{\jj\delta_j}}{s-M_j^2+\jj M_j\Gamma_j}
\right|^2\!,
\ee
where $M_j$ and $\Gamma_j$ are the mass and width of the $j$th intermediate
resonance, respectively. We will treat them as free parameters together 
with $R_j$ and $\delta_j$ (except $\delta_1$, which is kept at 0).

We first started fitting by assuming three resonances and got 
a very bad fit: $\cndf=137.5/99$, confidence level CL=0.6\% \cite{minuit}.
The participating resonances were $\phi(1020)$, $\phi(1680)$, and a wide
structure with mass of (1040$\pm$79)~MeV and $\Gamma=(1290\pm130)$~MeV 
representing the background. 

\begin{figure}[h]
\includegraphics[width=0.483\textwidth,height=0.5\textwidth]{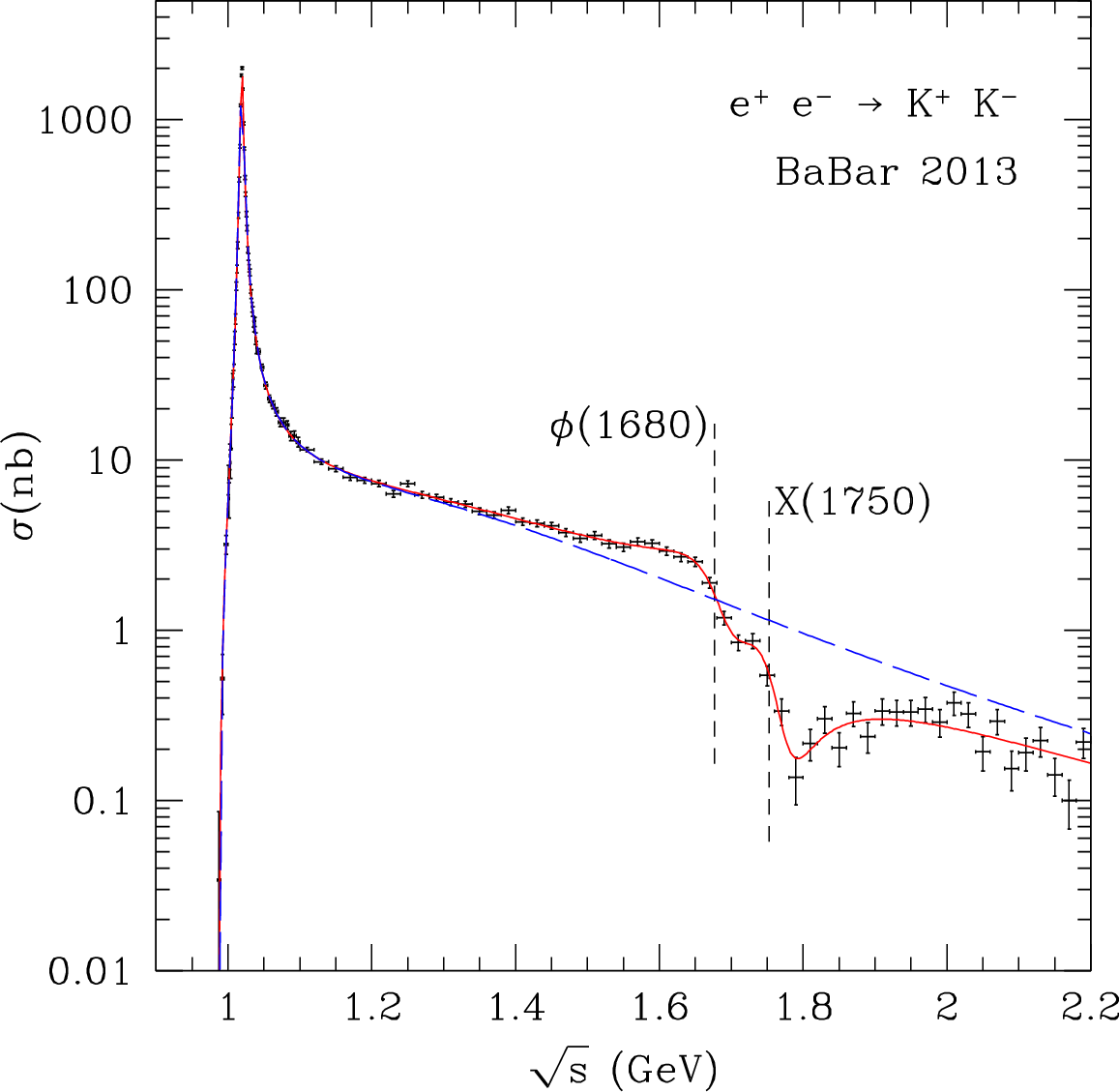}
\caption{\label{fig:babar2013}Fit to the \babar~data \cite{babar2013KpKm}
using Eq. \rf{sigma} with four resonances (full curve). The positions of
$\phi(1680)$ and X(1750) are marked with vertical lines. The model curve
without those two resonances is shown by the long dashes.}
\end{figure}

Then, we included a fourth resonance, and the fit quality improved 
to $\cndf=106.3/95$ (CL=20\%) \cite{foot1}. The parameters of all resonances 
are shown in Table \ref{tab:4reso}. The newcomer is the X(1750) resonance. 
\begin{table}[h]
\caption{\label{tab:4reso}Resonances appearing in the fit to the \babar~
data \cite{babar2013KpKm} using Eq. \rf{sigma} with $n\!=\!4$.}
\begin{tabular}{|l|ccc|}
\hline
 & $M$~(MeV) &  $\Gamma$~(MeV)& Significance \\
\hline
$\phi(1020)$ & 1019.072(16) & 4.284(35) & 87$\sigma$\\
$\phi(1680)$ & 1677$\pm$14  & 95$\pm$19 &  1.7$\sigma$\\
X(1750)      & 1753$\pm$21  & 77$\pm$18 &  1.0$\sigma$ \\
Background   & 1167$\pm$47  & 1040$\pm$89 & 4.9$\sigma$\\
\hline
\end{tabular}
\end{table}
The fit is compared with the data in Fig. \ref{fig:babar2013}. Despite
statistical significances of $\phi(1680)$ and X(1750) being low (1.7$\sigma$
and $1\sigma$, respectively), their effect on the shape of the excitation
curve between 1.5~GeV and 2~GeV is clearly visible. The reason is their
interference with the background, first constructive, then destructive.  
In order to appreciate the effect better, we show also the model prediction
with $\phi(1680)$ and X(1750) switched off.

We did not continue adding further resonances because our main aim was not
to get a perfect fit and find all resonances but to prove that the X(1750) 
is produced in the $\die\ra\dik$ process. The latter aim we have just 
accomplished.
\begin{figure}[t]
\includegraphics[width=0.483\textwidth,height=0.5\textwidth]{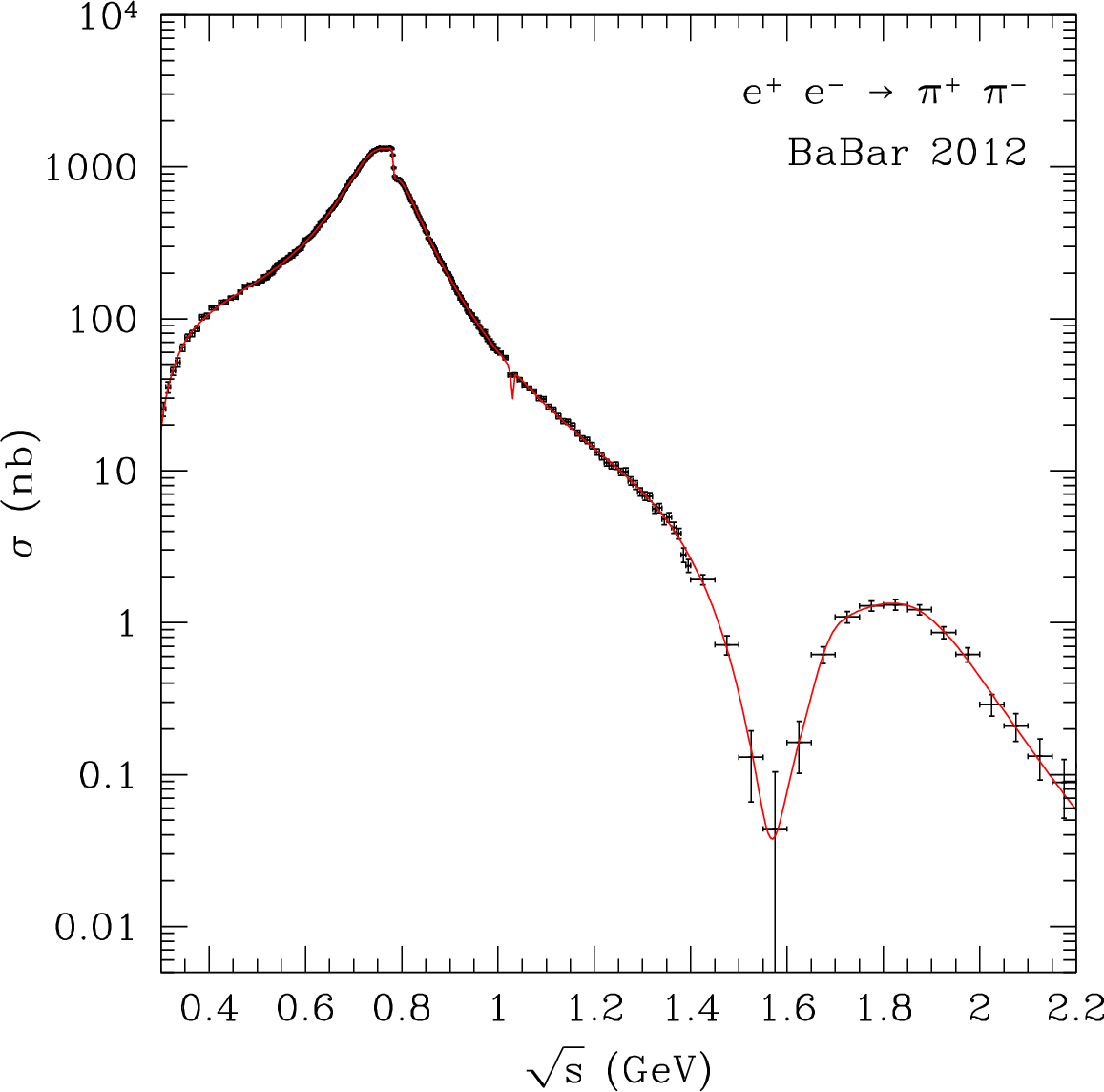}
\caption{\label{fig:babar2012}Fit to the \babar~data \cite{babar2012pippim}
using Eq. \rf{sigma} with six resonances. Only statistical errors were
considered. Quality of the fit is characterized by $\cndf=227.6/303$, 
CL=100\%.}
\end{figure}

In the next step, we explored very precise $\die\ra\pi^+\pi^-$ data of the 
\babar~Collaboration \cite{babar2012pippim} in the same energy range. 
We got a perfect fit with six resonances:
$\cndf=227.6/303$ (CL=100\%). With such a perfect fit, displayed in Fig.
\ref{fig:babar2012}, we can be almost sure that no resonance escaped our 
attention. 

The resonances found were: $\rho(770)$, $\omega(782)$,
$\rho(1450)$, and $\rho(1900)$. In addition, two resonances from the 
$\phi$ family: $\phi(1020)$ with statistical significance 0.2$\sigma$ and
$\phi(1680)$ with 0.3$\sigma$. We conclude that X(1750) production in the 
$\die\ra\pi^+\pi^-$ process is negligible.

The resonances of the $\phi$ family are characterized by the quantum numbers 
$I^G(J^{PC})=0^-(1^{--})$. What concerns the $\die$ annihilation is that 
they are copiously produced in the $\dik$ channel. However, in the 
$\pi^+\pi^-$ one, they appear at most at the level corresponding to the 
violation of the isospin ($G$ parity) symmetry \cite{phiinpippim}. That is 
precisely what the X(1750) does. Therefore, we suggest assigning isospin 
zero to it. 

This Letter shows another process, in addition to the photoproduction
and the decay of charmonium, where the X(1750) is produced. It hints 
that we can expect the participation of  X(1750) in all processes where 
its quantum numbers $I^G(J^{PC})=0^-(1^{--})$ permit.

The parameters of $\phi(1750)$ we found by analysing the $\die\ra\dik$
process, $M=(1753\pm21)$~MeV and  $\Gamma=(77\pm18)$~MeV, are compatible 
with those found in the other processes.

\acknowledgements
The author thanks Stanislav Hled\'{i}k and Josef Jur\'{a}\v{n} for the 
consultations about statistical aspects and Peter Lichard, Jr. for the 
assistance.

\end{document}